\documentclass{iau} 
\usepackage{graphicx}
\usepackage{latexsym}
\usepackage{natbib}
\usepackage{epsfig}
\def\mnras{MNRAS}
\def\aap{A\&A}
\def\apj{ApJ}
\def\apjl{ApJ}
\def\aj{AJ}

\def\pasp{PASP}
\def\apjs{ApJS}
\def\aaps{A\&AS}
\def\nat{Nature}

\def\oiiig{[O~{\small\ III}]$\,\lambda5007$}
\def\kms{km s$^{-1}$}
\title[PNe as dynamical tracers] 
{Planetary nebulae as  kinematic tracers of galaxy stellar
  halos}

\author[L. Coccato]   
{Lodovico Coccato}

\affiliation{European Southern Observatory, Karl-Schwarzschild Strasse 2, D-85748 Garching, Germany. email: {\tt lcoccato@eso.org}
}

\pubyear{2017}
\volume{323}  
\jname{Planetary nebulae: multiwavelength probes of stellar and galactic evolution.}
\editors{Xiaowei Liu, Letizia Stanghellini, \& Amanda Karakas, eds.}
\begin{document}

\maketitle

\begin{abstract}  
  The kinematic and dynamical properties of galaxy stellar halos are
  difficult to measure because of the faint surface brightness that
  characterizes these regions. Spiral galaxies can be probed using the
  radio H{\sc i} emission; on the contrary, early-type galaxies
  contain less gas, therefore alternative kinematic tracers need to be
  used. Planetary nebulae (PNe) can be easily detected far out in the
  halo thanks to their bright emission lines. It is therefore possible
  to map the halo kinematics also in early-type galaxies, typically
  out to 5 effective radii or beyond. Thanks to the recent
  spectroscopic surveys targeting extra-galactic PNe, we can now rely
  on a few tens of galaxies where the kinematics of the stellar halos
  are measured. Here, I will review the main results obtained in this field in the last
  decades.

\end{abstract}

\firstsection 
\section{Introduction}

The study of kinematics and chemical content in the outer regions of
galaxies is extremely important to understand dynamics, dark matter
content and distribution, and to get constrain on formation processes
of the galaxies themselves. Indeed, the dynamical timescale of the
stellar halos is comparable to the age of the galaxy (e.g.,
\citealt{Binney+87}), therefore the imprints of formation processes can
be still be preserved in the mass and orbital distribution and stellar
population properties.

The challenge of these studies resides in the faint surface brightness
level of the stellar halos. Indeed, either traditional long-slit or
integral field absorption line spectroscopy require long integration
times to go beyond 2 effective radii (e.g.,
\citealt{Coccato+10a,Coccato+10b,Coccato+11,
  Greene+12,Greene+15}). This is particularly challenging for
early-type galaxies because, contrarily to their spiral counterparts,
they do not contain as much gas to be used as kinematic tracer in
their outskirts.

Using planetary nebulae (PNe) as alternative tracer for the kinematics
of the underlying stellar population revealed to be a winning
strategy. Indeed, PNe can be detected far out in galaxy halos, thanks
to their bright \oiiig\ emission line that outshines the stellar
continuum background.
The idea has been exploited by many groups, targeting a large variety
of nearby galaxies including our own. The basic strategy includes the
identification of the PNe candidate (e.g., by means of multi band
photometry) and their spectroscopic follow-up to get the radial velocity
and, eventually, emission line ratios for chemical studies.
Because this two-stages approach can include a number of systematic
effects (e.g. change of observing conditions between exposures), a
dedicated instrument was build: Planetary Nebulae Spectrograph (PN.S,
\citealt{Douglas+02}). The PN.S exploits the simultaneous
counter-dispersed imaging technique, which allows to identify and
obtain the radial velocity of a PNe in a single exposure.

The purpose of this review is to highlight the main results in
obtained by using PNe as kinematic tracers in the last decades.

\section{Measuring the distribution and kinematics of the PNe population}
\label{sec:analysis}

The first thing that needs to be done to establish if PNe are indeed
useful to extend the information of stars to larger radii is  to
check whether or not the PNe represent the properties of the
underlying stellar population in the regions where their observations
overlap.
This is done in two ways: by comparing the number density of PNe with
the stellar surface brightness profiles (Section \ref{sec:sb}), and by
comparing the stellar kinematics with the one derived from absorption
line spectroscopy (Section \ref{sec:kinem}).

\subsection{Spatial distribution of PNe}
\label{sec:sb}

The most common way to compare the PNe density distribution and
stellar surface brightness is divide the galaxy into concentric
annuli, count the PNe within each annuls, and compare the number
density (counts divided by area) to the mean stellar surface
brightness in that area. One important aspect to consider is the
completeness of the sample, which can be established by populating the
original image with simulated sources and re-run the PNe detection
procedure. The number density of PNe needs to be corrected for
incompleteness before being compared to the stellar surface
brightness. The scaling factor between light and PNe counts is related
to the PNe specific frequency (the so-called $\alpha$ parameter,
\citealt{Jacoby80}); see for example \citet{Arnaboldi12} for a review.

\subsection{Kinematics}
\label{sec:kinem}

 Unlike absorption line spectroscopy,
in which the kinematics measured at a given position {\it is} the luminosity
weighted kinematics of all the stars along that line of sight, the
velocity of a given PN represents that object only, and does not
necessarily reflect the mean behaviour of the population. In other
words, a single PNe velocity determination is like a ``random number''
extracted from the line-of-sight velocity distribution.
Thus, it is necessary to proceed on a statistical way, and use the
combined information of many PNe around a given area to
``reconstruct'' the mean kinematics (velocity and velocity dispersion)
of that area. Then, one can compare the kinematics obtained from stars
and PNe in the regions where both tracers are available.

\begin{figure}
\psfig{file=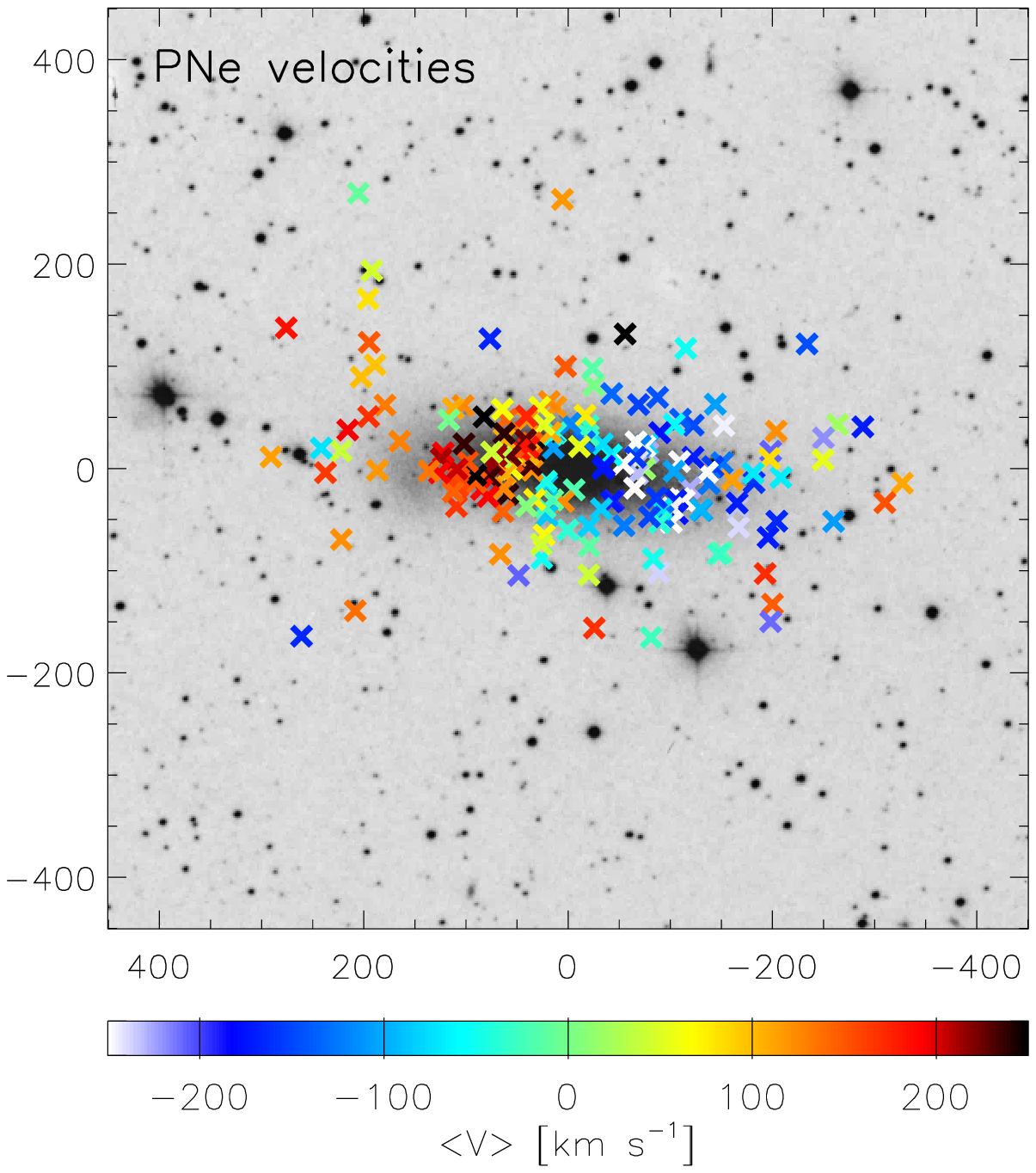,width=4.45cm, bb=51  336 409 730, clip=}
\psfig{file=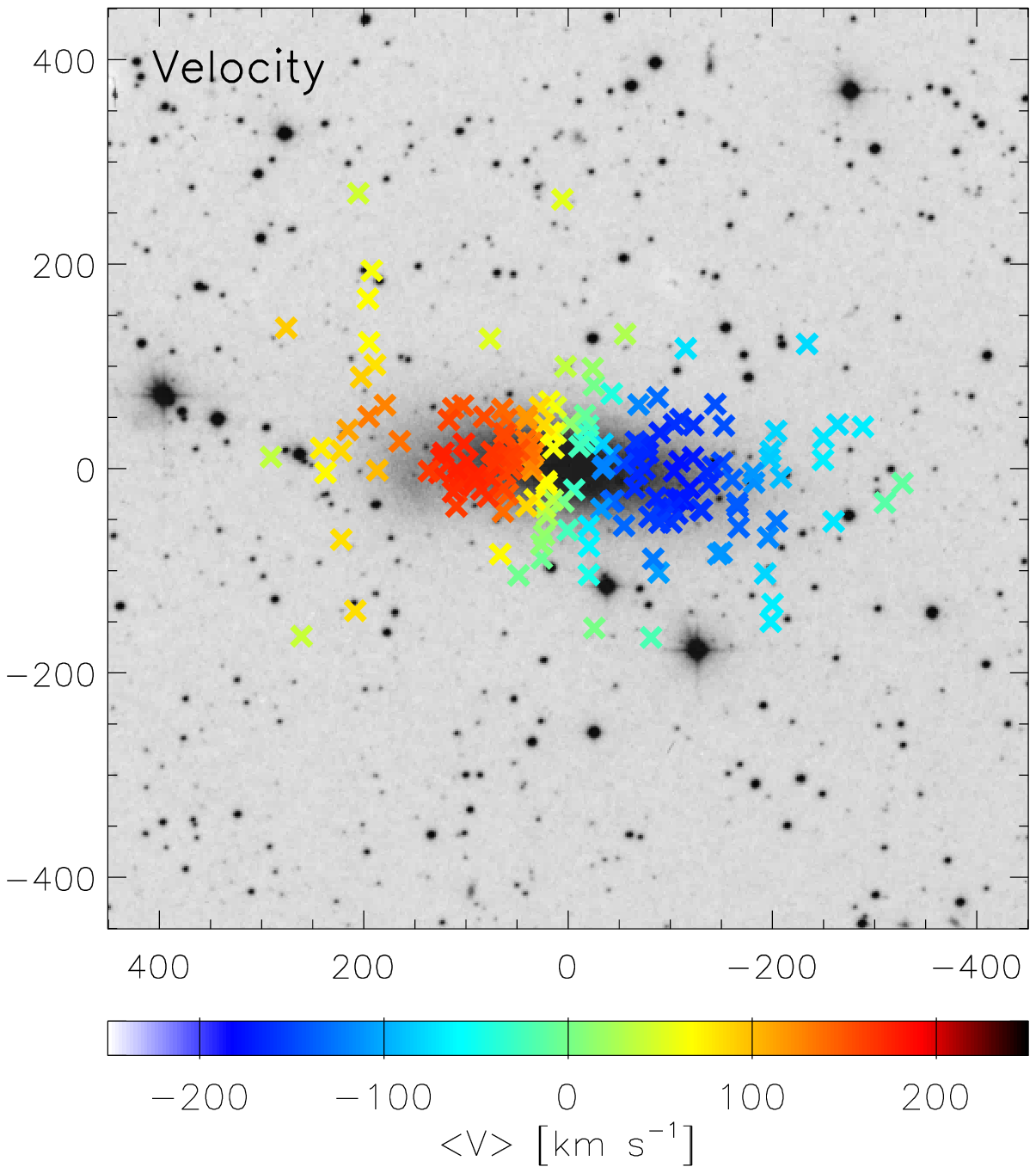,width=4.45cm, bb=51  336 409 730, clip=}
\psfig{file=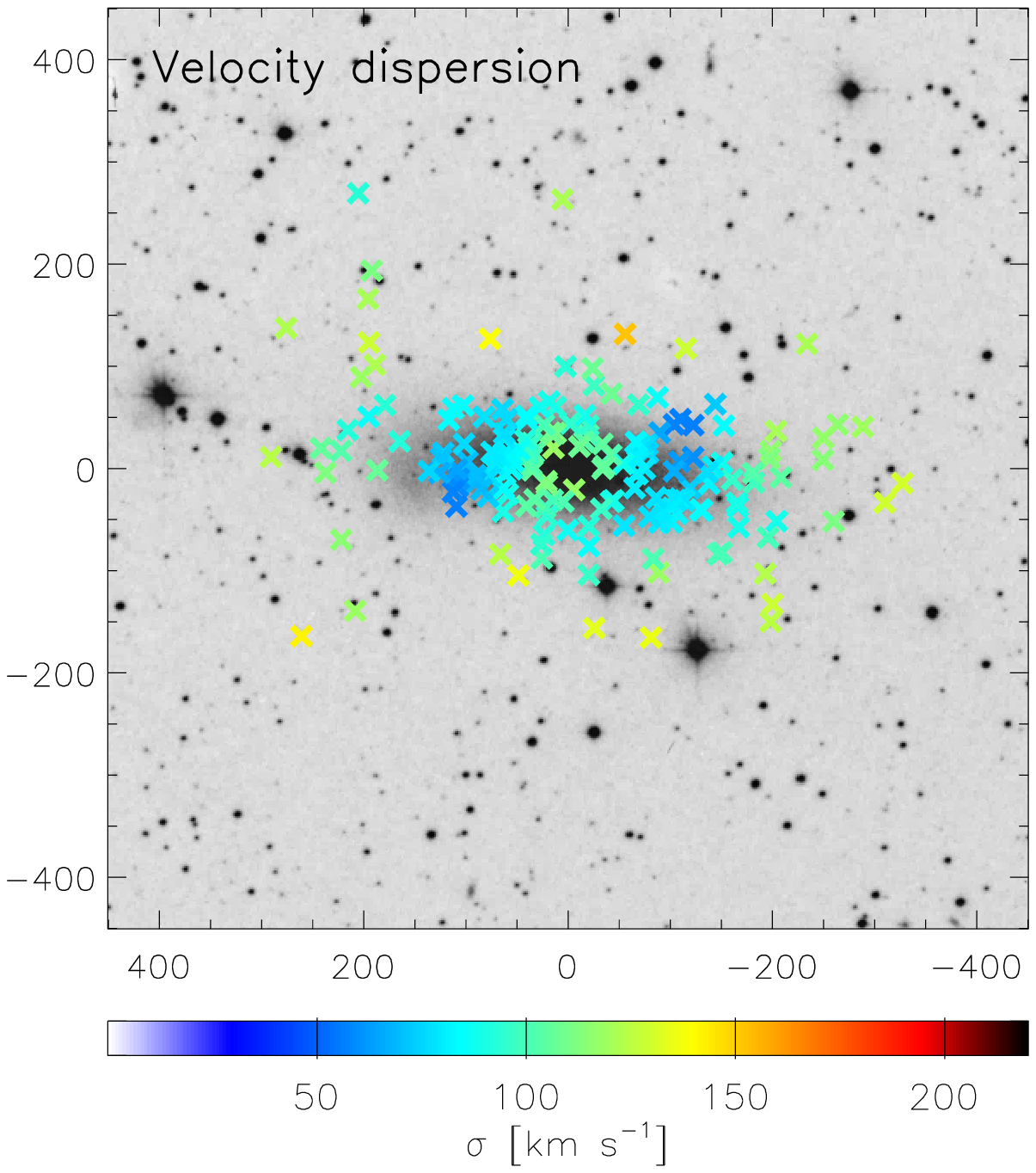,width=4.45cm, bb=51  336 409 730, clip=}
\caption{Recovering the kinematics of the PNe population from
  individual radial velocities. The left panel shows the position of
  the PNe (crosses) detected in NGC 1023, superimposed to the image of
  the field of view (data from \citealt{Noordermeer+08}). The color of
  each cross is proportional to the radial velocity of the PNe (minus
  the galaxy systemic velocity), as indicated by the color bar. The
  central and right panels show the values of the reconstructed
  velocity and velocity dispersion fields at each PNe position, using
  the adaptive Gaussian kernel smoothing. The spatial scale is in arcseconds.}
\label{fig:2dfield} 
\end{figure}

Several techniques have been developed either to reconstruct the 2D
velocity kinematics or to use the individual velocities in dynamical
models \citep{Romanowsky+01, Amorisco+12, Foster+13, Arnold+14}. Here
we focus on the ``Gaussian Kernel smoothing'' technique developed in
\citet{Peng+04}, with ``Adaptive Kernel'' as introudced in
\citet{Coccato+09}. Basically, the kinematics at each point in the
field of view is obtained computing weighted moments of the velocity
distribution of the entire PNe population. Each PN contributes with a
weight that depends on its distance to that point. The weights
dependency from distance is a Gaussian function, where its dispersion
(kernel amplitude) depends on the local density of PNe. Regions with
high PNe density have smaller kernel amplitude, regions with low PNe
density have larger kernel amplitude.
Figure \ref{fig:2dfield} compares the two-dimensional map of
individual PN velocities with the two-dimensional maps of velocity and
velocity dispersions reconstructed using this technique.
 
\subsection{Comparison between stellar and PNe properties}

The comparison between stars and PNe revealed that the distribution
and kinematics of PNe is in good agreement with the stellar surface
brightness and kinematics in the regions where these tracers
overlap. Moreover, for early-type galaxies, the PNe spatial
distribution is consistent with the extrapolation of the S\'ersic fit
to the stellar surface brightness, rather than a de Vaucouleurs plus an
exponential disk, in line with the view of ``spheroidal'' halos around
early-type galaxies, rather than ``disk-like'' structures
(e.g. \citealt{Coccato+09}, and references therein).

These finding have an important practical application: the combined
information of the kinematics and distribution of stars (that
typically probe the central effective radius) and PNe (in the outer
regions) can be used to derive the entire dynamic picture of a galaxy
and its halo (see Section \ref{sec:mass}).

\begin{figure}
\begin{center}
\psfig{file=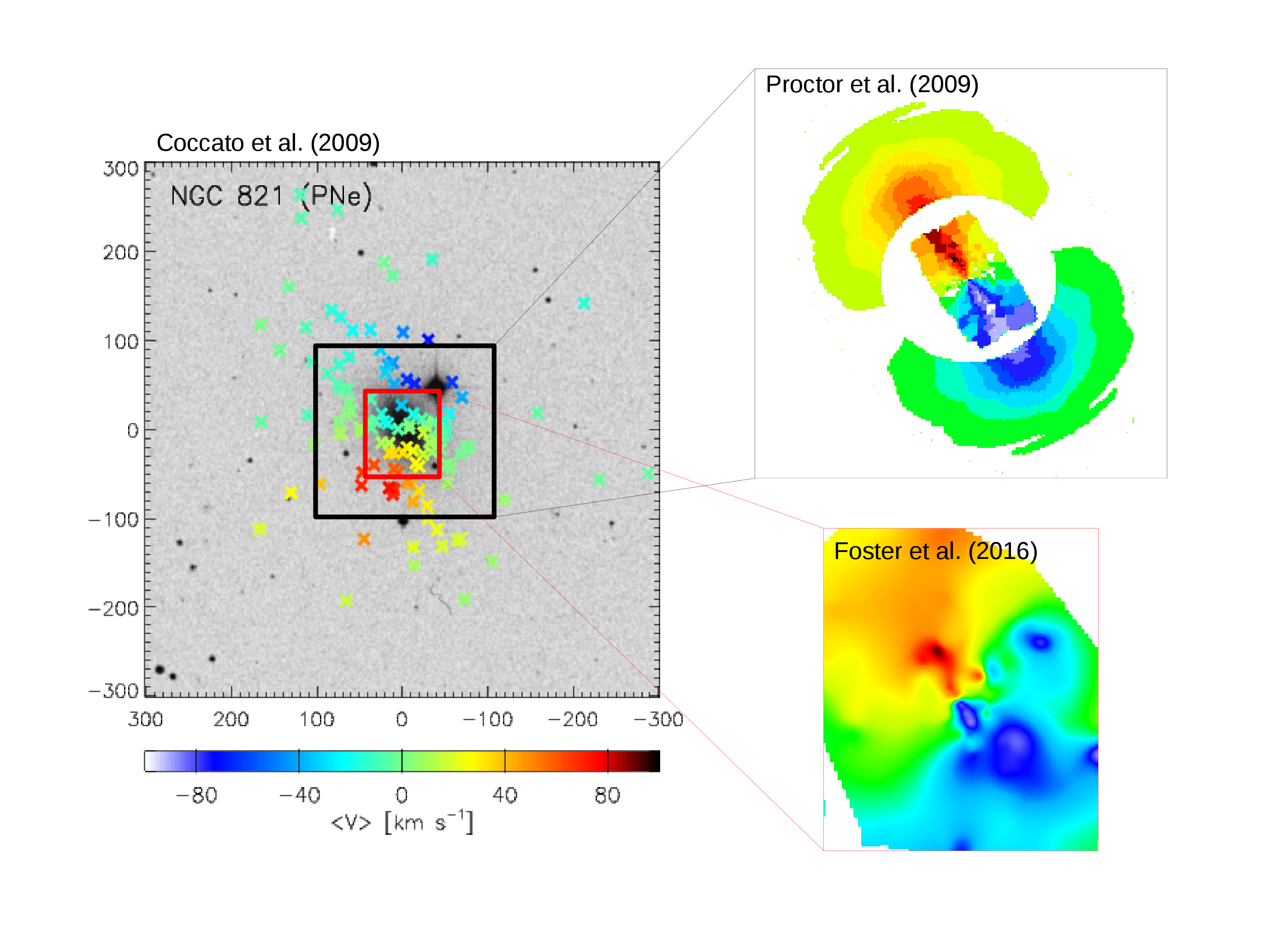,width=8.6cm, bb=44 49 735 557, clip=}
\end{center}
\caption{Comparison between the velocity field of the PNe population
  (left panel) compared with the velocity field of the stars (right
  panels).  PNe data are from \citet{Coccato+09}, stellar data are
  from \citet[for the inner $\sim 200''\times200"$]{Proctor+09} and
  \citet[for the inner $\sim 100"\times120''$]{Foster+16}. The
  direction of rotation of PNe is $\sim 100^{\circ}$ misaligned with
  respect to the direction of rotation of the and stars. The spatial
  scale on the left panel is in arcseconds.}
\label{fig:821} 
\end{figure}

There are, however, few exceptions to this  picture. The most
outstanding example is given by E6 galaxy NGC 821 (Figure \ref{fig:821}). Its PNe system
\citep{Coccato+09} shows a rotation axis which is $\sim 100^{\circ}$
misaligned from that of the stars \citep{Proctor+09, Foster+16}. These
exceptions and, more in general, the differences between the
kinematics obtained from different tracers, are seen as an indication
of recent satellite accretion in the halos \citep{Coccato+13,
   Foster+14, Romanowsky+14, Longobardi+15}.

\section{Results}

\subsection{Early works}
\subsubsection{The Milky-Way}
The use of PNe radial velocities to measure the kinematic properties
of a galaxy was first applied to our own Milky Way. In their
pioneering work, \citet{Schneider+83} exploited the measured radial
velocities of 252 galactic PNe to derive the galactic rotation curve
beyond the solar circle. Their results showed a rising rotation curve
out to $\sim 15$ kpc, in agreement and extending further out, the
previous results obtained with O-B stars and CO clouds.

The study was extended also to the MW bulge by \citet{Durand+98} using
867 galactic PNe and later by \citet{Beaulieu+00} using 373 bulge
PNe. At odd with previous results, there was no evidence for a rising
disk rotation curve in these studies. The kinematic of the bulge was
found to be consistent with a linearly increasing rotation velocity
\citep{Durand+98}. The comparison with N-body models showed that the
kinematics of the bulge is consistent with models where the MW bulge
and bar formed from the disk via bar-forming instabilities
\citep{Beaulieu+00}.

Nowadays, the census of galactic PNe with radial velocities includes
about 2500 objects (e.g., the HASH database, \citealt{Parker+16}).

\subsubsection{The Andromeda Galaxy M31}

The next obvious candidate for these studies was our neighbor
galaxy M31. \citet{Nolthenius+87} detected and measured the radial
velocities of 37 PNe within 15 and 30 kpc along the photometric major
axis of M31. Despite the scatter in their measurements, they showed
that the rotation curve of M31 flattens at $\sim 220$ \kms\ outside 20
kpc from the galaxy center.  The first quantum step in the number of
PNe detection in M31 and its satellites was done in
\citet{Merrett+06}, with the publication of a catalog of $\sim 2700$
sources. This catalog allowed to identify a new satellite (Andromeda
VIII) and to probe the stellar distribution in M31 out to large
distance from the galaxy center: they found is no indication of a
cut-off in M31's disk out to four scale lengths, and no signs of a
spheroidal halo population in excess of the bulge out to 10 effective
bulge radii.

\subsection{Early-Types}

\begin{figure}
\begin{center}
\psfig{file=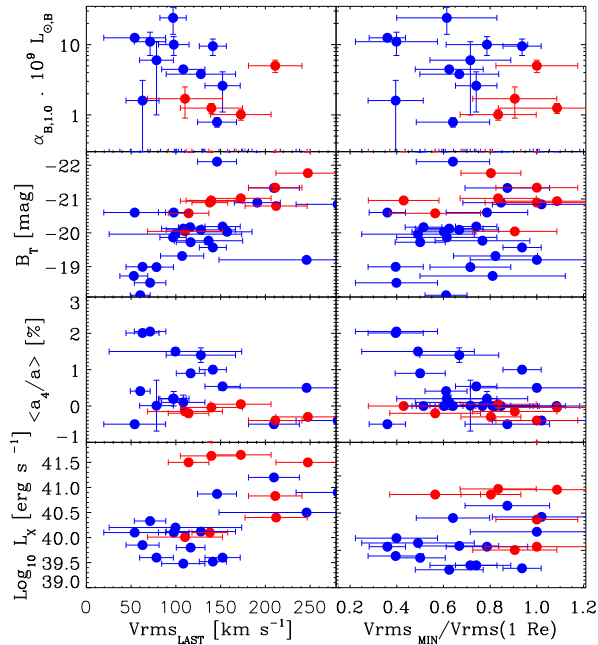,width=8.cm, clip=}
\end{center}
\caption{Correlations between the halo kinematics and other galaxy
  properties (updated from \citealt{Coccato+09}). The halo kinematics
  is parametrized with the outermost value of $V_{\rm rms}$ (left) and its
  normalization at 1 effective radius (right), as obtained from PNe
  radial velocities. Each point represents a galaxy. From top to
  bottom, the panels indicate: PNe specific frequency, total B-band
  luminosity, shape parameter, and total X-ray luminosity. Galaxies
  that are classified as fast rotators are shown in blue, slow
  rotators in red.}
\label{fig:correlations} 
\end{figure}

The use of PNe radial velocity measurements has revealed the following
kinematic properties for the halos of early-type galaxies (e.g.,
\citealt{Douglas+07, Coccato+09, Teodorescu+10, McNeil+10}, and references
therein):

\begin{itemize}
\item The velocity dispersion and $V_{\rm RMS}$ have a large variety
  of radial profiles, including flat profiles (generally associated to
  more massive galaxies), declining profiles (generally associated to
  less massive ellipticals), and even rising profiles (generally
  associated to lenticulars with a prominent stellar halo, or
  disturbed systems).

\item Kinematic twists and misalignment in the velocity fields are
  more frequent at large radii than what observed in the inner
  regions. Twists are also observed in some fast rotator galaxies.

\item Outer halos are characterized by more complex radial profiles
  of the specific angular momentum-related $\lambda(R)$ parameter than
  observed within one effective radius. Some galaxies are more
  rotational dominated at large radii than in their central parts
  (i.e., the $\lambda(R)$ profile increase in the halo), some other
  are pressure supported at large radii (i.e., the $\lambda(R)$
  profile decreases in the halo).

\item The halo kinematics are correlated with other galaxy properties,
  such as total B-band and X-ray luminosity, isophotal shape, total
  stellar mass, $V/\sigma$ and PNe specific frequency, with a clear
  separation between fast and slow rotators (Figure
  \ref{fig:correlations}).

\item That the halos of massive ellipticals sometimes host kinematic
  substructures, interpreted as relics of accretion events. Kinematic
  substructures are  also found in the halos of spiral galaxies (e.g.,
  \citealt{Herrmann+09a}).
\end{itemize}


\subsubsection{Mass distribution in Early-Type galaxies: ``lost and found'' Dark Matter}
\label{sec:mass}

The main science driver for studying the kinematics of PNe in
Early-Type galaxies is to use them as tracer for the gravitational
potential in order to infer the mass distribution and the fraction of
dark matter.

Initial works that included a significant amount of PNe in the
analysis were carried on massive nearby ellipticals, such as NGC 1316
\citep{Arnaboldi+98}, NGC 5128 \citep{Hui+95, Peng+04}, and more
recently on NGC 4649 \citep{Teodorescu+11} - just to cite few
examples. The signature of the presence of dark matter was clearly
present in the kinematic profile at large radii.
Other works led by several team that combined the information of other
tracers to the PNe, reached similar results (e.g.,
\citealt{Napolitano+05, Deason+12, Forbes+16}).

Controversial results emerged from the study of ordinary
ellipticals. Studies by \citet{Ciardullo+93, Mendez+01, Mendez+09,
  Romanowsky+03} indicated that the kinematic profile
of intermediate luminosity ellipticals are consistent with no dark
matter, or much less that what predicted by the $\Lambda$CDM model.
The different teams adopted different modeling techniques, including
isotropic or constant anisotropic Jeans models fitting several moments
of the PNe velocity distribution, which lead to results consistent to
each others.
These findings challenged not only the view that {\it all} galaxies
are embedded in massive halos, but also alternative theories such as MOND
\citep{Milgrom83}, that were not able to reconcile the luminous mass
in these galaxies with the observed decreasing velocity dispersion
profiles.

The solution was found by assuming that the stellar orbits are
anisotropic, and that their anisotropy varies with radius (becoming
more radial outward). This reconciled the observed dispersion profiles
with MOND \citep{Milgrom+03}. Moreover the increase of radial
anisotropy at large radii is consistent with the numerical simulations
of disk-galaxy mergers in a $\Lambda$CDM universe \citep{Dekel+05}.

\begin{figure}
\begin{center}
\psfig{file=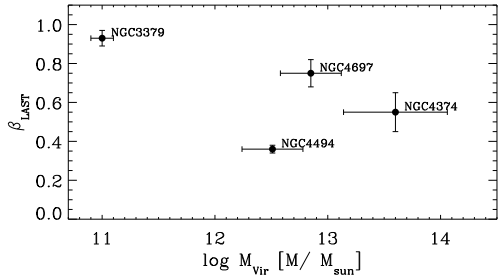,width=7.5cm, clip=}
\end{center}
\caption{Trend between virial mass (from \citealt{Forbes+17}, Table 1)
  and anisotropy parameter $\beta$ (from Fig. 23 in
  \citealt{Morganti+13}, and from Fig. A2 in
  \citealt{Napolitano+11}). The value of $\beta$ refers to the
  outermost determination. Radially biased orbits are those with high
  values of $\beta$; isotropic models have $\beta=0$.}
\label{fig:mass_beta} 
\end{figure}

Following this idea, more elaborated dynamical models of PNe datasets
revealed that the inclusion of radial variation in the anisotropy
profile was indeed able to reconcile the dark matter content in
low-luminous ellipticals with the predictions of $\Lambda$CDM, altough
a certan amount of degeneracy between mass, shape, and anisotropy
exists for nearly spherical systems (e.g., \citealt{DeLorenzi+09}). In
addition, as indicated in Figure \ref{fig:mass_beta}, the new models
suggest that lower mass halos have stars in orbits that have higher
radial anisotropy than the stars in high mass halos
(\citealt{Morganti+13}, \citealt{Napolitano+11}, Napolitano,
priv. comm.), in line with what found also using other tracers (e.g.,
\citealt{Zhu+16}).

\subsection{Lenticulars}
Lenticular galaxies are mainly composed by a spheroid and a disk
without spiral arms. Therefore, it is reasonable to assume that some
PNe are associated to the spheroid and some to the disk. The
association of a PNe to one component or the other can be done by
combining the surface brightness spheroid/disk decomposition with PNe
kinematic data, assuming a non rotating model for the spheroid
component \citep{Cortesi+11}.  One application is to recover the disk
kinematics by removing the contamination from the PNe associated to
the spheroid in order to study scaling relations such as
the Tully-Fisher at large radii.  \citet{Cortesi+13b} found that: i)
the disks of S0s have higher velocity dispersion than the spirals with
the same rotational velocity; and that ii) S0s lie around one
magnitude below the Tully-Fisher relation for spiral galaxies, i.e.,
disks of S0s have higher circular velocity that those of spirals of
the same luminosity. This was interpreted by the authors as an
indication of a mild harassment process that that heated disks of
spirals and transformed them into S0s.

\subsection{Spirals}

Altough spiral galaxies have gas that can be used to probe their
kinematics and mass distribution at large radii, PNe are still useful
to probe the vertical stellar velocity dispersion ($\sigma_Z$) in
nearly face-on systems. \citet{Herrmann+09b} studied a sample of five
nearby, low-inclination spiral galaxies. They found that: i) 4/5
spirals have constant mass-to-light ratio out to $\sim 3$ optical
scale lengths ($h_Z$); ii) outside $\sim 3 h_Z$, $\sigma_Z$ becomes
flat with radius; iii) the disks of early type spirals have higher
values of mass-to-light ratio and are closer to maximal than the disks
of later-type spirals; iv) the inner halos of spirals are better fit
by pseudo-isothermal laws than by NFW models.

The PNe in spiral galaxies were also used to measure the stellar
migration in stellar disks to investigate its influence on the
variation of metallicity gradients in stellar disks with time
\citep{Magrini+16}.

\section{Summary}

In this review I have illustrated the main findings obtained by using
planetary nebulae as kinematic tracers in galaxy halos, sometimes
persuing an historical approach. The main lessons that we can learn in
these last decades on the topics are:
\begin{itemize}
\item In the regions were PNe and absorption line data overlap, PNe
  well trace the surface brightness and kinematics of stars (with few
  exceptions). Therefore, PNe can be used to measure the
  properties of galaxy halos by extending the information of stellar
  kinematics to large radii, and to unveil signatures of accretions
  events also in galaxies with a relatively undisturbed morphology.
\item Variable radial anisotropy needs to be considered in dynamical
  models in order to obtain a reliable estimate of the dark matter
  content and distribution in the halos. There is the indication lower
  mass halos have stars in orbits that have higher radial anisotropy
  than the stars in high mass halos.
\item The properties of the PNe population in spiral galaxies suggest
  that the disks of early type spirals have higher values of
  mass-to-light ratio and are closer to maximal than the disks of
  later-type spirals.
\item The study of PNe kinematics in S0 galaxies revealed that diks of
  S0s have higher velocity dispersion that those of spirals, and that
  S0s lie one magnitude below the Tully-Fisher relation with respect
  to spiral galaxies of the same luminosity.
\end{itemize}


\begin{thebibliography}{54}
\expandafter\ifx\csname natexlab\endcsname\relax\def\natexlab#1{#1}\fi

\bibitem[{{Amorisco} \& {Evans}(2012)}]{Amorisco+12}
{Amorisco}, N.~C. \& {Evans}, N.~W. 2012, \mnras, 424, 1899

\bibitem[{{Arnaboldi}(2012)}]{Arnaboldi12}
{Arnaboldi}, M. 2012, in IAU Symposium, Vol. 283, IAU Symposium, 267--274

\bibitem[{{Arnaboldi} {et~al.}(1998){Arnaboldi}, {Freeman}, {Gerhard},
  {Matthias}, {Kudritzki}, {M{\'e}ndez}, {Capaccioli}, \&
  {Ford}}]{Arnaboldi+98}
{Arnaboldi}, M., {Freeman}, K.~C., {Gerhard}, O., {et~al.} 1998, \apj, 507, 759

\bibitem[{{Arnold} {et~al.}(2014){Arnold}, {Romanowsky}, {Brodie}, {Forbes},
  {Strader}, {Spitler}, {Foster}, {Blom}, {Kartha}, {Pastorello}, {Pota},
  {Usher}, \& {Woodley}}]{Arnold+14}
{Arnold}, J.~A., {Romanowsky}, A.~J., {Brodie}, J.~P., {et~al.} 2014, \apj,
  791, 80

\bibitem[{{Beaulieu} {et~al.}(2000){Beaulieu}, {Freeman}, {Kalnajs}, {Saha}, \&
  {Zhao}}]{Beaulieu+00}
{Beaulieu}, S.~F., {Freeman}, K.~C., {Kalnajs}, A.~J., {Saha}, P., \& {Zhao},
  H. 2000, \aj, 120, 855

\bibitem[{{Binney} \& {Tremaine}(1987)}]{Binney+87}
{Binney}, J. \& {Tremaine}, S. 1987, {Galactic dynamics}

\bibitem[{{Ciardullo} {et~al.}(1993){Ciardullo}, {Jacoby}, \&
  {Dejonghe}}]{Ciardullo+93}
{Ciardullo}, R., {Jacoby}, G.~H., \& {Dejonghe}, H.~B. 1993, \apj, 414, 454

\bibitem[{{Coccato} {et~al.}(2013){Coccato}, {Arnaboldi}, \&
  {Gerhard}}]{Coccato+13}
{Coccato}, L., {Arnaboldi}, M., \& {Gerhard}, O. 2013, \mnras, 436, 1322

\bibitem[{{Coccato} {et~al.}(2010{\natexlab{a}}){Coccato}, {Arnaboldi},
  {Gerhard}, {Freeman}, {Ventimiglia}, \& {Yasuda}}]{Coccato+10a}
{Coccato}, L., {Arnaboldi}, M., {Gerhard}, O., {et~al.} 2010{\natexlab{a}},
  \aap, 519, A95

\bibitem[{{Coccato} {et~al.}(2010{\natexlab{b}}){Coccato}, {Gerhard}, \&
  {Arnaboldi}}]{Coccato+10b}
{Coccato}, L., {Gerhard}, O., \& {Arnaboldi}, M. 2010{\natexlab{b}}, \mnras,
  407, L26

\bibitem[{{Coccato} {et~al.}(2009){Coccato}, {Gerhard}, {Arnaboldi}, {Das},
  {Douglas}, {Kuijken}, {Merrifield}, {Napolitano}, {Noordermeer},
  {Romanowsky}, {Capaccioli}, {Cortesi}, {de Lorenzi}, \&
  {Freeman}}]{Coccato+09}
{Coccato}, L., {Gerhard}, O., {Arnaboldi}, M., {et~al.} 2009, \mnras, 394, 1249

\bibitem[{{Coccato} {et~al.}(2011){Coccato}, {Gerhard}, {Arnaboldi}, \&
  {Ventimiglia}}]{Coccato+11}
{Coccato}, L., {Gerhard}, O., {Arnaboldi}, M., \& {Ventimiglia}, G. 2011, \aap,
  533, A138

\bibitem[{{Cortesi} {et~al.}(2011){Cortesi}, {Merrifield}, {Arnaboldi},
  {Gerhard}, {Martinez-Valpuesta}, {Saha}, {Coccato}, {Bamford}, {Napolitano},
  {Das}, {Douglas}, {Romanowsky}, {Kuijken}, {Capaccioli}, \&
  {Freeman}}]{Cortesi+11}
{Cortesi}, A., {Merrifield}, M.~R., {Arnaboldi}, M., {et~al.} 2011, \mnras,
  414, 642

\bibitem[{{Cortesi} {et~al.}(2013){Cortesi}, {Merrifield}, {Coccato},
  {Arnaboldi}, {Gerhard}, {Bamford}, {Napolitano}, {Romanowsky}, {Douglas},
  {Kuijken}, {Capaccioli}, {Freeman}, {Saha}, \& {Chies-Santos}}]{Cortesi+13b}
{Cortesi}, A., {Merrifield}, M.~R., {Coccato}, L., {et~al.} 2013, \mnras, 432,
  1010

\bibitem[{{de Lorenzi} {et~al.}(2009){de Lorenzi}, {Gerhard}, {Coccato},
  {Arnaboldi}, {Capaccioli}, {Douglas}, {Freeman}, {Kuijken}, {Merrifield},
  {Napolitano}, {Noordermeer}, {Romanowsky}, \& {Debattista}}]{DeLorenzi+09}
{de Lorenzi}, F., {Gerhard}, O., {Coccato}, L., {et~al.} 2009, \mnras, 395, 76

\bibitem[{{Deason} {et~al.}(2012){Deason}, {Belokurov}, {Evans}, \&
  {McCarthy}}]{Deason+12}
{Deason}, A.~J., {Belokurov}, V., {Evans}, N.~W., \& {McCarthy}, I.~G. 2012,
  \apj, 748, 2

\bibitem[{{Dekel} {et~al.}(2005){Dekel}, {Stoehr}, {Mamon}, {Cox}, {Novak}, \&
  {Primack}}]{Dekel+05}
{Dekel}, A., {Stoehr}, F., {Mamon}, G.~A., {et~al.} 2005, \nat, 437, 707

\bibitem[{{Douglas} {et~al.}(2002){Douglas}, {Arnaboldi}, {Freeman}, {Kuijken},
  {Merrifield}, {Romanowsky}, {Taylor}, {Capaccioli}, {Axelrod}, {Gilmozzi},
  {Hart}, {Bloxham}, \& {Jones}}]{Douglas+02}
{Douglas}, N.~G., {Arnaboldi}, M., {Freeman}, K.~C., {et~al.} 2002, \pasp, 114,
  1234

\bibitem[{{Douglas} {et~al.}(2007){Douglas}, {Napolitano}, {Romanowsky},
  {Coccato}, {Kuijken}, {Merrifield}, {Arnaboldi}, {Gerhard}, {Freeman},
  {Merrett}, {Noordermeer}, \& {Capaccioli}}]{Douglas+07}
{Douglas}, N.~G., {Napolitano}, N.~R., {Romanowsky}, A.~J., {et~al.} 2007,
  \apj, 664, 257

\bibitem[{{Durand} {et~al.}(1998){Durand}, {Acker}, \& {Zijlstra}}]{Durand+98}
{Durand}, S., {Acker}, A., \& {Zijlstra}, A. 1998, \aaps, 132, 13

\bibitem[{{Forbes} {et~al.}(2016){Forbes}, {Alabi}, {Romanowsky}, {Brodie},
  {Strader}, {Usher}, \& {Pota}}]{Forbes+16}
{Forbes}, D.~A., {Alabi}, A., {Romanowsky}, A.~J., {et~al.} 2016, \mnras, 458,
  L44

\bibitem[{{Forbes} {et~al.}(2017){Forbes}, {Alabi}, {Romanowsky}, {Kim},
  {Brodie}, \& {Fabbiano}}]{Forbes+17}
{Forbes}, D.~A., {Alabi}, A., {Romanowsky}, A.~J., {et~al.} 2017, \mnras, 464,
  L26

\bibitem[{{Foster} {et~al.}(2013){Foster}, {Arnold}, {Forbes}, {Pastorello},
  {Romanowsky}, {Spitler}, {Strader}, \& {Brodie}}]{Foster+13}
{Foster}, C., {Arnold}, J.~A., {Forbes}, D.~A., {et~al.} 2013, \mnras, 435,
  3587

\bibitem[{{Foster} {et~al.}(2014){Foster}, {Lux}, {Romanowsky},
  {Mart{\'{\i}}nez-Delgado}, {Zibetti}, {Arnold}, {Brodie}, {Ciardullo},
  {GaBany}, {Merrifield}, {Singh}, \& {Strader}}]{Foster+14}
{Foster}, C., {Lux}, H., {Romanowsky}, A.~J., {et~al.} 2014, \mnras, 442, 3544

\bibitem[{{Foster} {et~al.}(2016){Foster}, {Pastorello}, {Roediger}, {Brodie},
  {Forbes}, {Kartha}, {Pota}, {Romanowsky}, {Spitler}, {Strader}, {Usher}, \&
  {Arnold}}]{Foster+16}
{Foster}, C., {Pastorello}, N., {Roediger}, J., {et~al.} 2016, \mnras, 457, 147

\bibitem[{{Greene} {et~al.}(2015){Greene}, {Janish}, {Ma}, {McConnell},
  {Blakeslee}, {Thomas}, \& {Murphy}}]{Greene+15}
{Greene}, J.~E., {Janish}, R., {Ma}, C.-P., {et~al.} 2015, \apj, 807, 11

\bibitem[{{Greene} {et~al.}(2012){Greene}, {Murphy}, {Comerford}, {Gebhardt},
  \& {Adams}}]{Greene+12}
{Greene}, J.~E., {Murphy}, J.~D., {Comerford}, J.~M., {Gebhardt}, K., \&
  {Adams}, J.~J. 2012, \apj, 750, 32

\bibitem[{{Herrmann} \& {Ciardullo}(2009)}]{Herrmann+09b}
{Herrmann}, K.~A. \& {Ciardullo}, R. 2009, \apj, 705, 1686

\bibitem[{{Herrmann} {et~al.}(2009){Herrmann}, {Ciardullo}, \&
  {Sigurdsson}}]{Herrmann+09a}
{Herrmann}, K.~A., {Ciardullo}, R., \& {Sigurdsson}, S. 2009, \apjl, 693, L19

\bibitem[{{Hui} {et~al.}(1995){Hui}, {Ford}, {Freeman}, \& {Dopita}}]{Hui+95}
{Hui}, X., {Ford}, H.~C., {Freeman}, K.~C., \& {Dopita}, M.~A. 1995, \apj, 449,
  592

\bibitem[{{Jacoby}(1980)}]{Jacoby80}
{Jacoby}, G.~H. 1980, \apjs, 42, 1

\bibitem[{{Longobardi} {et~al.}(2015){Longobardi}, {Arnaboldi}, {Gerhard}, \&
  {Mihos}}]{Longobardi+15}
{Longobardi}, A., {Arnaboldi}, M., {Gerhard}, O., \& {Mihos}, J.~C. 2015, \aap,
  579, L3

\bibitem[{{Magrini} {et~al.}(2016){Magrini}, {Coccato}, {Stanghellini},
  {Casasola}, \& {Galli}}]{Magrini+16}
{Magrini}, L., {Coccato}, L., {Stanghellini}, L., {Casasola}, V., \& {Galli},
  D. 2016, \aap, 588, A91

\bibitem[{{McNeil} {et~al.}(2010){McNeil}, {Arnaboldi}, {Freeman}, {Gerhard},
  {Coccato}, \& {Das}}]{McNeil+10}
{McNeil}, E.~K., {Arnaboldi}, M., {Freeman}, K.~C., {et~al.} 2010, \aap, 518,
  A44

\bibitem[{{M{\'e}ndez} {et~al.}(2001){M{\'e}ndez}, {Riffeser}, {Kudritzki},
  {Matthias}, {Freeman}, {Arnaboldi}, {Capaccioli}, \& {Gerhard}}]{Mendez+01}
{M{\'e}ndez}, R.~H., {Riffeser}, A., {Kudritzki}, R.-P., {et~al.} 2001, \apj,
  563, 135

\bibitem[{{M{\'e}ndez} {et~al.}(2009){M{\'e}ndez}, {Teodorescu}, {Kudritzki},
  \& {Burkert}}]{Mendez+09}
{M{\'e}ndez}, R.~H., {Teodorescu}, A.~M., {Kudritzki}, R.-P., \& {Burkert}, A.
  2009, \apj, 691, 228

\bibitem[{{Merrett} {et~al.}(2006){Merrett}, {Merrifield}, {Douglas},
  {Kuijken}, {Romanowsky}, {Napolitano}, {Arnaboldi}, {Capaccioli}, {Freeman},
  {Gerhard}, {Coccato}, {Carter}, {Evans}, {Wilkinson}, {Halliday}, \&
  {Bridges}}]{Merrett+06}
{Merrett}, H.~R., {Merrifield}, M.~R., {Douglas}, N.~G., {et~al.} 2006, \mnras,
  369, 120

\bibitem[{{Milgrom}(1983)}]{Milgrom83}
{Milgrom}, M. 1983, \apj, 270, 365

\bibitem[{{Milgrom} \& {Sanders}(2003)}]{Milgrom+03}
{Milgrom}, M. \& {Sanders}, R.~H. 2003, \apjl, 599, L25

\bibitem[{{Morganti} {et~al.}(2013){Morganti}, {Gerhard}, {Coccato},
  {Martinez-Valpuesta}, \& {Arnaboldi}}]{Morganti+13}
{Morganti}, L., {Gerhard}, O., {Coccato}, L., {Martinez-Valpuesta}, I., \&
  {Arnaboldi}, M. 2013, \mnras, 431, 3570

\bibitem[{{Napolitano} {et~al.}(2005){Napolitano}, {Capaccioli}, {Romanowsky},
  {Douglas}, {Merrifield}, {Kuijken}, {Arnaboldi}, {Gerhard}, \&
  {Freeman}}]{Napolitano+05}
{Napolitano}, N.~R., {Capaccioli}, M., {Romanowsky}, A.~J., {et~al.} 2005,
  \mnras, 357, 691

\bibitem[{{Napolitano} {et~al.}(2011){Napolitano}, {Romanowsky}, {Capaccioli},
  {Douglas}, {Arnaboldi}, {Coccato}, {Gerhard}, {Kuijken}, {Merrifield},
  {Bamford}, {Cortesi}, {Das}, \& {Freeman}}]{Napolitano+11}
{Napolitano}, N.~R., {Romanowsky}, A.~J., {Capaccioli}, M., {et~al.} 2011,
  \mnras, 411, 2035

\bibitem[{{Nolthenius} \& {Ford}(1987)}]{Nolthenius+87}
{Nolthenius}, R. \& {Ford}, H.~C. 1987, \apj, 317, 62

\bibitem[{{Noordermeer} {et~al.}(2008){Noordermeer}, {Merrifield}, {Coccato},
  {Arnaboldi}, {Capaccioli}, {Douglas}, {Freeman}, {Gerhard}, {Kuijken}, {de
  Lorenzi}, {Napolitano}, \& {Romanowsky}}]{Noordermeer+08}
{Noordermeer}, E., {Merrifield}, M.~R., {Coccato}, L., {et~al.} 2008, \mnras,
  384, 943

\bibitem[{{Parker} {et~al.}(2016){Parker}, {Boji{\v c}i{\'c}}, \&
  {Frew}}]{Parker+16}
{Parker}, Q.~A., {Boji{\v c}i{\'c}}, I.~S., \& {Frew}, D.~J. 2016, Journal of
  Physics Conference Series, 728, 032008

\bibitem[{{Peng} {et~al.}(2004){Peng}, {Ford}, \& {Freeman}}]{Peng+04}
{Peng}, E.~W., {Ford}, H.~C., \& {Freeman}, K.~C. 2004, \apj, 602, 685

\bibitem[{{Proctor} {et~al.}(2009){Proctor}, {Forbes}, {Romanowsky}, {Brodie},
  {Strader}, {Spolaor}, {Mendel}, \& {Spitler}}]{Proctor+09}
{Proctor}, R.~N., {Forbes}, D.~A., {Romanowsky}, A.~J., {et~al.} 2009, \mnras,
  398, 91

\bibitem[{{Romanowsky} {et~al.}(2014){Romanowsky}, {Arnold}, {Brodie},
  {Foster}, {Forbes}, {Lux}, {Mart{\'{\i}}nez-Delgado}, {Strader}, {Zibetti},
  \& {Sluggs Team}}]{Romanowsky+14}
{Romanowsky}, A.~J., {Arnold}, J.~A., {Brodie}, J.~P., {et~al.} 2014, in
  Astronomical Society of the Pacific Conference Series, Vol. 486, Multi-Spin
  Galaxies, ASP Conference Series, ed. E.~{Iodice} \& E.~M. {Corsini}, 169

\bibitem[{{Romanowsky} {et~al.}(2003){Romanowsky}, {Douglas}, {Arnaboldi},
  {Kuijken}, {Merrifield}, {Napolitano}, {Capaccioli}, \&
  {Freeman}}]{Romanowsky+03}
{Romanowsky}, A.~J., {Douglas}, N.~G., {Arnaboldi}, M., {et~al.} 2003, Science,
  301, 1696

\bibitem[{{Romanowsky} \& {Kochanek}(2001)}]{Romanowsky+01}
{Romanowsky}, A.~J. \& {Kochanek}, C.~S. 2001, \apj, 553, 722

\bibitem[{{Schneider} \& {Terzian}(1983)}]{Schneider+83}
{Schneider}, S.~E. \& {Terzian}, Y. 1983, \apjl, 274, L61

\bibitem[{{Teodorescu} {et~al.}(2010){Teodorescu}, {M{\'e}ndez}, {Bernardi},
  {Riffeser}, \& {Kudritzki}}]{Teodorescu+10}
{Teodorescu}, A.~M., {M{\'e}ndez}, R.~H., {Bernardi}, F., {Riffeser}, A., \&
  {Kudritzki}, R.~P. 2010, \apj, 721, 369

\bibitem[{{Teodorescu} {et~al.}(2011){Teodorescu}, {M{\'e}ndez}, {Bernardi},
  {Thomas}, {Das}, \& {Gerhard}}]{Teodorescu+11}
{Teodorescu}, A.~M., {M{\'e}ndez}, R.~H., {Bernardi}, F., {et~al.} 2011, \apj,
  736, 65

\bibitem[{{Zhu} {et~al.}(2016){Zhu}, {Romanowsky}, {van de Ven}, {Long},
  {Watkins}, {Pota}, {Napolitano}, {Forbes}, {Brodie}, \& {Foster}}]{Zhu+16}
{Zhu}, L., {Romanowsky}, A.~J., {van de Ven}, G., {et~al.} 2016, \mnras, 462,
  4001

\end{thebibliography}

\end{document}